\documentclass[5p,times]{elsarticle}

\usepackage{lineno,hyperref}

\usepackage{url}
\usepackage{changepage}
\usepackage{algorithm}
\usepackage{multirow}
\usepackage{epsfig}
\usepackage[noend]{algpseudocode}
\usepackage[usenames, dvipsnames]{color}
\usepackage{subfig}
\usepackage{caption}
\usepackage{enumitem}
\usepackage{array}

\algnewcommand{\Inputs}[1]{%
  \Statex \textbf{Inputs:}
  \Statex \hspace*{\algorithmicindent}\parbox[t]{.8\linewidth}{\raggedright #1}
}
\algnewcommand{\Outputs}[1]{%
  \Statex \textbf{Outputs:}
  \Statex \hspace*{\algorithmicindent}\parbox[t]{.8\linewidth}{\raggedright #1}
}
\algnewcommand{\Initialize}[1]{%
  \State \textbf{Initialize:}
  \Statex \hspace*{\algorithmicindent}\parbox[t]{.8\linewidth}{\raggedright #1}
}

\newcolumntype{M}[1]{>{\centering\arraybackslash}m{#1}}

\modulolinenumbers[5]

\journal{Journal of Information and Software Technology}

\begin{document}
\begin{frontmatter}

\title{Incorporating fault-proneness estimations into coverage-based test case prioritization methods}

\author[sharifaddress]{Mostafa Mahdieh}

\author[sharifaddress]{Seyed-Hassan Mirian-Hosseinabadi
\corref{mycorrespondingauthor}}
\cortext[mycorrespondingauthor]{Corresponding author}
\ead{hmirian@sharif.edu}
\author[sharifaddress]{Khashayar Etemadi}
\author[sharifaddress]{Ali Nosrati}
\author[sharifaddress]{Sajad Jalali}

\address[sharifaddress]{Sharif University of Technology}

\begin{abstract}
\textit{Context:} During the development process of a software program, regression testing is used to ensure that the correct behavior of the software is retained after updates to the source code. This regression testing becomes costly over time as the number of test cases increases and it makes sense to prioritize test cases in order to execute fault-detecting test cases as soon as possible. There are many coverage-based test case prioritization (TCP) methods that only use the code coverage data to prioritize test cases. By incorporating the fault-proneness estimations of code units into the coverage-based TCP methods, we can improve such techniques.

\textit{Objective:} In this paper, we aim to propose an approach which improves coverage-based TCP methods by considering the fault-proneness distribution over code units. Further, we present the results of an empirical study that shows using our proposed approach significantly improves the additional strategy, which is a widely used coverage-based TCP method.

\textit{Method:} The approach presented in this study uses the bug history of the software in order to introduce a defect prediction method to learn a neural network model. This model is then used to estimate fault-proneness of each area of the source code and then the estimations are incorporated into coverage-based TCP methods. Our proposed approach is a general idea that can be applied to many coverage-based methods, such as the additional and total TCP methods.

\textit{Results:} The proposed methods are evaluated on datasets collected from the development history of five real-world projects including 357 versions in total. The experiments show that using an appropriate bug history can improve coverage-based TCP methods.

\textit{Conclusion:} The proposed approach can be applied to various coverage-based TCP methods and the experiments show that it can improve these methods by incorporating estimations of code units fault-proneness.

\end{abstract}


\end{frontmatter}


\section{Introduction}
Regression testing is a testing activity meant to assure that updates to the software have not changed the existing software behavior. Regression test suites normally grow in size as the software is developed or more quality assurance activity is performed. Although creating more test cases help for the test suite to be more effective, a large test suite is costly to execute and its execution might take hours or even days to finish.

Test case prioritization (TCP) seeks to help testers by prioritizing test cases in an order that testers gain maximum benefit from. For this purpose, test cases are prioritized and executed in an order that optimizes a desired goal function. The common target used for TCP is to minimize the time needed for finding failed test cases. Finding failed test cases earlier helps the development team to work on issues and resolve them faster in the development process. 

The vast majority of TCP methods use structural coverage as a metric to prioritize test cases \cite{yoo2012regression, hao2014unified}. Coverage-based TCP methods usually aim to put test cases in an order that reaches a high coverage as soon as possible. There are two major strategies in coverage-based TCP methods, namely, {\it total } and {\it additional } strategies \cite{rothermel1999test}, which we will introduce in Section~\ref{coverage_based_tcp}. Recent studies have shown that even an optimal coverage-based TCP method does not perform much better than the additional method in terms of fault detection rate \cite{hao2016optimal}. This suggests that in order to improve the additional method some sources of data other than the structural coverage data should be used.

Another source of information that can be used for TCP is the fault-proneness information. A high fault-proneness for a code unit indicates it is relatively probable that the code unit contains a fault. There are various approaches for leveraging fault-proneness information to improve TCP; hence, there is a challenge to choose an appropriate approach in this regard. In this paper, we propose a novel approach to incorporate fault-proneness information into coverage-based TCP methods. Furthermore, in order to estimate the fault-proneness of a code unit, we designed a novel neural network based defect prediction method customized for the current problem. Our approach is generally applicable to many coverage-based TCP methods. In this study, we will specifically apply it to the total and additional TCP methods to obtain \textit{modified total} and \textit{modified additional} methods. Our experiment on 357 versions of five real-world projects, included in the Defects4J dataset \cite{just2014defects4j}, shows that the fault detection rate of these methods is improved using the mentioned modification.

To assess our research, we ask the following research questions:

\begin{itemize}[label={}]
    \item {\bf RQ1:} How does the modified additional TCP strategy compare to the traditional additional TCP strategy in terms of fault detection performance?
    \item {\bf RQ2:} How does the modified total TCP strategy compare to the traditional total TCP strategy in terms of fault detection performance?
    \item {\bf RQ3:} How does the value of the modification tuning parameters (explained in Section \ref{modified_coverage}) affect the effectiveness of the modified strategies?
\end{itemize}

This paper makes the following contributions:
\begin{itemize}
\item We propose a novel method to incorporate fault-proneness estimations into coverage-based TCP methods. Our approach is based on a modification of the concept of coverage in coverage-based TCP.
\item
We design a customized neural network based defect prediction method to estimate the fault-proneness of a code unit. This method is customized to work when only a small set of bugs is available and utilizes the information from all versions of the source code history.
\item We present an extension of the Defects4J dataset. For each program version included in Defects4J, our extended dataset contains the test coverage data and computed values of some source code metrics of that version. The source code metric data is used by the developed defect prediction method.
\item
We present an empirical evaluation using five open-source projects containing in total 357 versions of the projects. Results show that our proposed modification could improve existing coverage-based TCP techniques.

\end{itemize}

The rest of the paper is organized as follows: Section~\ref{background_sec} presents the background material.  Section~\ref{proposed_method_sec} presents our approach of solving the problem and our proposed method. Section~\ref{empirical_study_sec} presents the setup of our empirical evaluation and Section~\ref{section_results} shows the results of our experiments. In Section~\ref{section_discussion} the empirical results and threats to the validity of this study are discussed. Section~\ref{related_work_sec} summarizes the most related work to this paper. Finally, Section~\ref{section_conclusion} contains the conclusions and future work of this paper.

\section{Background}
\label{background_sec}
In this section, we present the definition of test case prioritization and briefly introduce coverage-based test case prioritization methods. We continue by providing background information on defect prediction, which is related to our study.

\subsection{Test case prioritization}
Consider a test suite containing the set of test cases $T = \{ t_1, t_2, \ldots, t_n \}$. The TCP problem is formally defined as follows \cite{elbaum2002test}:

\begin{adjustwidth}{1cm}{}
{\it Given:} $T$, a test suite; $PT$, the set of permutations of T; $f$, a function from $PT$ to the real numbers.\\	
{\it Problem:} Find $T' \in PT$ such that\footnote{This relation is expressed using Z notation's first order logic \cite{usingz}.}:
\begin{equation}
\label{prioritization_problem_1} 
\forall T'' : PT \mid T'' \ne T' \bullet f(T') \ge f(T'')
\end{equation}

\end{adjustwidth}

In other words, the problem of TCP is to find a permutation $T'$ such that $f(T')$ is maximum. Here $f$ is a scoring function that assuming a permutation from $PT$, assigns a score value to that permutation.

Users of TCP methods could have different goals, such as testing high-risk components as soon as possible, maximizing code coverage, and detecting faults at a faster rate. The $f$ function represents the goal of a TCP activity.

The APFD (Average Percentage of Faults Detected) goal function, an area-under-curve metric that measures how quickly a test suite can detect faults, is frequently used in the literature for TCP when the goal of TCP is maximizing the fault detection rate \cite{rothermel1999test, yoo2012regression, engstrom2011improving, catal2013test}. The other target function that can be used is the percentage of test cases executed until the first failing test case. The first-failing test case metric is also applicable in our case, however the APFD function is more commonly used in the field of test case prioritization, and using it enables the results to be comparable to future studies of TCP.
APFD is defined as follows: Denote the set of all failed test cases in $T'$ as $T_f = \langle t_{f_1}, t_{f_2}, \ldots, t_{f_l} \rangle $, where $l$ is the number of failed test cases and the index of test case $t_{f_i}$ in $T'$ is $f_i$. The APFD target function is formulated as \cite{rothermel2001prioritizing}:
\begin{equation}
APFD = 1 - \frac{\sum\limits_{i=1}^l f_i}{nl} + \frac{1}{2n}
\end{equation}

For a permutation in which the failed test cases are executed earlier, the values of $f_1$ to $f_l$ are smaller, so the APFD value will be larger.

\subsection{Coverage-based test case prioritization}
\label{coverage_based_tcp}
In order to measure coverage, the source code is partitioned into hierarchical units such as packages, files, methods and statements. Coverage-based TCP methods choose one level of partitioning (usually statements or methods) and define coverage over those units. Assuming a chosen level of partitioning, consider the source partitioned into units $U = \{u_1, u_2, \ldots, u_m \}$.

For each test case $t_i$ and unit $u_j$ of the code, $Cover(i, j)$ denotes whether test case $t_i$ covers unit $u_j$. The amount of coverage is either 0 or 1 if the units of code are statements; however, it can also be a real number in the range $[0,1]$ if the units are methods, files, classes or packages.

The total coverage of a test case $t_i$ is usually defined as follows:
\begin{equation}
\label{total_coverage_equation}
Cover(i) = \sum\limits_{1 \le j \le m} Cover(i,j)
\end{equation}

The basic idea of the \textit{total prioritization strategy} is that test cases with more coverage have more chance to uncover bugs. The total strategy therefore, sorts test cases by the amount of source code that each test case covers. This strategy ignores the fact that some test cases might cover the same area of the code. Therefore, when test cases are sorted using this strategy, usually some units of code are run multiple times, before the whole units are covered \cite{elbaum2002test}.

On the other hand, the \textit{additional prioritization strategy} assumes that running an uncovered unit of the code has more priority compared to an already covered unit. The intuition behind the additional strategy is that early coverage of all units of the code, results in revealing faults sooner \cite{elbaum2002test}.

Coverage information can be collected in two ways. Dynamic coverage information is collected by executing the program and tracking every unit that is executed. In contrast, static coverage is derived by static analysis on the source code \cite{mei2012static}. We apply dynamic coverage in our study, as it is generally more accurate than static coverage and usually leads to more effective prioritization results.

\subsection{Defect Prediction}
It is frequently observed that some areas of the code are relatively more fault-prone (i.e., more defects occur in that areas throughout the development of software). This happens due to various features of some code areas, such as relative complexity of the implementation, more code churn, and faulty designs. 

Issue trackers and bug databases contain important information about software failures. This wealth of information can be analyzed by defect prediction methods to identify fault-prone areas of the code. Defect prediction applies machine learning to analyze the bug history of software and produce a prediction model for fault-proneness.

Defect prediction methods often include the following conceptual steps \cite{nam2014survey}:
\begin{enumerate}
\item{Feature extraction}: In this step, metrics from the code and development process and other sources of information are extracted as a feature vector for each unit of the code (package, file, class or method). Moreover, the number of previous bugs related to each unit of the code is extracted and stored.
\item{Model learning}: Data extracted from the previous step is fed to a machine learning or data mining algorithm to learn a prediction model. While creating this prediction model, the metrics extracted in the previous step are used as the feature vector and the number of previous bugs related to each code unit is used as the target function.
\item{Validation/Prediction}:
The model learned from the previous step can now be used to assign each unit of code a fault-proneness score. Some part of the data (namely, the validation set) is usually withdrawn from the learning procedure. The validation set is used to validate the prediction strength of the model.
\end{enumerate}

There are many studies using static code metrics for defect prediction \cite{menzies2010defect,menzies2007data,zimmermann2007predicting}. Other metrics, such as historical and process related metrics (e.g., number of past bugs \cite{klas2010transparent,ostrand2005predicting} or number of changes \cite{hassan2005top, pinzger2008can, meneely2008predicting, moser2008comparative}) and organizational metrics (e.g., number of developers \cite{weyuker2008too, graves2000predicting}), have also been used.

\section{Methodology}
\label{proposed_method_sec}
In this section, we introduce our proposed approach for TCP. Since our approach is a modification of existing TCP methods, before introducing our approach, we review the existing methods that we evaluate their modified versions in our empirical study.

\subsection{Review of traditional random, total, and additional TCP methods}
\label{traditional_strategies}
In this subsection, we review three traditional TCP strategies that we compare them with their modified strategies in our empirical study.

\subsubsection{Random strategy}
Due to the random strategy, test cases are randomly sorted. The APFD of this strategy is about $50\%$ and it is usually used as a baseline to be compared with other proposed strategies for evaluation \cite{ashraf2012value, elbaum2004selecting}.

\subsubsection{Total strategy for TCP}
The total strategy for TCP begins by computing the total coverage of all test cases according to Equation~\ref{total_coverage_equation}. In the next step, test cases are sorted due to their total coverage so that the first test case has the highest total coverage. Compared with other non-random existing strategies, this strategy is simple, efficient. The time complexity of the total algorithm consists of the time complexity of computing the total coverage for all test cases plus the time complexity of the used sorting algorithm. The summation of these items results in the time complexity of the total algorithm that is $\mathcal{O}(nm + n\log{}n)$.

\subsubsection{Additional strategy for TCP}
\label{trad_add_strat}
The same as the total strategy, the additional strategy begins by computing the total coverage of all test cases. Afterwards, a greedy algorithm is used to prioritize the test cases. Due to this algorithm, in each step, the test case that has the highest coverage over the uncovered code area is chosen as the next test case. The selected test case is then appended to the end of the ordered list of test cases and marked not to be chosen in next steps. Moreover, the area of the code covered by this chosen test case will be marked as covered area.

This strategy works in $n$ steps where $n$ shows the number of test cases. In each step, selecting the next test case and updating the coverage of the remaining test cases is done in $\mathcal{O}(nm)$ where computing the updated coverage of a test case is performed in $\mathcal{O}(m)$. Therefore, the total time complexity of this algorithm is $\mathcal{O}(n^2 m)$.

The additional strategy can be implemented in different variations. In two situations, this strategy faces different options:

\begin{itemize}
    \item When more than one non-selected test cases have the highest coverage over the uncovered code area: in this case, one of these test cases should be selected with some criteria. For example, one might select the test case randomly or select the next test case with higher coverage over the whole code area (i.e. covered and uncovered).
    \item When all areas of the code are covered by the test cases that have already been selected: in this case, the coverage of all the remaining test cases would be $0\%$. Again, the remaining test cases can be ordered with different criteria. For instance, one might order them randomly or due to their total coverage. Another option is to consider all the code uncovered and repeat the algorithm with remaining test cases again \cite{elbaum2002test}. 
\end{itemize}

\subsection{Proposed approach}
In this subsection, we describe our proposed approach and the rationale behind it.

\subsubsection{Motivation}
We have two main motivations for proposing our approach: First, previous studies related to defect prediction have suggested that defect prediction methods be leveraged for automated tasks, such as test case prioritization \cite{lewis2013does}. Second, developers usually tend to firstly test those parts of the program that are more likely to be faulty; however, existing TCP methods generally do not consider this tendency of developers. By incorporating the fault-proneness score, estimated by learned defect prediction models, into TCP methods we address these two mentioned motivations.

\subsubsection{Modified coverage}
\label{modified_coverage}
Assuming that we can extract prior knowledge on the fault probability of the code units, we propose a modified coverage formula that incorporates this prior knowledge. Given that the probability of existing faults in unit $u_j$ ($1 \le j \le m$) is $Prob(F_j)$\footnote{$F_j$ indicates the event in which \textit{j}th code unit is faulty and $Prob(F_J)$ represents the probability of this event.}, we propose the following modified coverage formula to compute the coverage for the test case $t_i$:

\begin{equation}
\label{weighted_cover}
FaultBasedCover(i) = \sum\limits_{1 \le j \le m} Cover(i,j) \times Prob(F_j)
\end{equation}

This formula considers more weight for units with more fault probability, resulting in giving more priority to these units.

To utilize this modified coverage formula, we need to estimate a probability function that represents the estimation of the probability of a defect existing in each unit of the source code. In this manner, we designed an appropriate defect prediction method, which we present in Section \ref{proposed_dp_method}. The defect prediction method assigns a fault-proneness score $P_{dp}(j)$ to every unit $u_j$ in the code ($1 \le j \le m$). In order to incorporate the fault-proneness score into the traditional TCP methods, one option is to define the probability function $Prob(F_j)$ as follows:

\begin{equation}
\label{prob_as_fault_proneness}
    Prob(F_j) = P_{dp}(j)
\end{equation}

Using Equation~\ref{prob_as_fault_proneness} as the definition of the probability function has the implication that covering the code units that are not predicted to be fault-prone will completely be ignored. Nonetheless, this is a downside for this definition because our prior knowledge indicates that even parts of the code that are not predicted to be fault-prone might contain some faults.

In contrast with Equation~\ref{prob_as_fault_proneness}, we can also define the probability function as the following equation which leads to ignoring the fault-proneness of unit codes:

\begin{equation}
\label{probability_defined_as_constant}
    Prob(F_j) = 1
\end{equation}

In order to avoid ignoring the fault-proneness of unit codes or the test coverage over the code units that are not predicted to be fault-prone, we introduce the following definition for the probability function:

\begin{equation}
\label{proposed_probability_function}
Prob(F_j) = P_0 + (1-P_0) \times P_{dp}(j)
\end{equation}

In Equation~\ref{proposed_probability_function}, $P_0$ should be a real number between $0$ and $1$. If it is set to $1$, the result would be the same as Equation~\ref{prob_as_fault_proneness} and if it is set to $0$, the result would be similar to Equation~\ref{probability_defined_as_constant}.

\subsubsection{Proposed defect prediction method}
\label{proposed_dp_method}
Defect prediction methods work at various granularity levels \cite{zimmermann2007predicting, hata2012bug, kamei2013large}. Nevertheless, mainstream research on defect prediction has been mainly focused on file level defect prediction. Therefore, we designed a file/class level defect prediction method for regression test prioritization. We use the fault-proneness score of the classes as an estimation for the fault proneness score of the methods.

\begin{figure}
\centering
\includegraphics[width=3.6in]{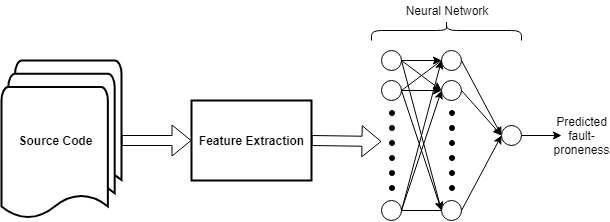}
\caption{Overview of the defect prediction method.}
\label{defect_prediction_method}
\end{figure}

Figure~\ref{defect_prediction_method} illustrates our defect prediction method. In this method, each file of the source code that had a bug in a past version was marked as buggy. A feature extractor was designed to extract source code and historical features related to bug prediction. The feature extractor is then executed on each class of the source code, resulting in a feature vector for each class. The details about the extracted features will be discussed further in Section~\ref{empirical_study_sec}.

The next step for bug prediction is to learn a prediction model relating the source code feature vectors and the fault-proneness of the class. For this purpose, a neural network with two layers was learned with the following structure:
\begin{itemize}
\item The first layer (i.e. input layer) has an input neuron for each of the features extracted by the feature extractor. The total number of input features is the sum of all values of the column \textit{Feature count} in Table~\ref{tab:features} (except the bug label), which is 104. Therefore, this layer contains 104 neurons.
\item The second layer (i.e. hidden layer) has 300 neurons with a sigmoid activation function.
\item The final output neuron has the sigmoid activation function.
\end{itemize}
The neural network is trained to solve a binary classification problem with two classes of \textit{fault-prone} and \textit{not-fault-prone}. Although all the class values in the training phase are either zero (not-fault-prone) or one (fault-prone), the neural network learns to output  a real number between zero and one i.e. in the range of $[0,1]$. For each input source code class, this number can be interpreted as the fault-proneness score of the class.

Various classification methods have been proposed for defection prediction \cite{lessmann2008benchmarking}. Certain challenges are raised due to the specific differences that our defect prediction problem has compared to a usual one. These differences can be summarized as follows:
\begin{enumerate}
\item Traditional defect prediction methods work with the assumption that source code classes do not change too much over time \cite{menzies2010defect}. Hence, they only consider the feature vector extracted from the last source code version and count the total number of bugs in previous versions for each class. The mentioned assumption leads to errors in the learning process and because of the role that defect prediction plays in our approach, we need to minimize such errors in our study.
\item The size of our set of positive samples (the number of classes marked as buggy) is small (the size and other properties of the dataset used in our study will be further discussed in Section~\ref{subjects_of_study}).
\end{enumerate}


To consider the first difference between our defect prediction problem and a usual one, we design our method to take into account the data from all versions of the source code for each class. In this manner, we put together all the feature vectors of all versions of the source code, resulting in a larger training set, approximately with the size $\sum\limits_{v \in Versions} |Classes(v)|$.

This would cause the dataset to be more unbalanced, compared to when a single feature vector is extracted for each class, because the size of the positive samples would be nearly the same while the number of negative samples would be multiplied by a factor of $|Versions|$. Moreover, the dataset would include some similar feature vectors with inconsistent labels. This happens because a non-buggy class can become buggy with a few changes; therefore, the dataset will include two very similar feature vectors related to this class with different labels. We apply a neural network learning method which is appropriate for this situation.

To address the second difference between our defect prediction problem and a usual one, we propose learning the neural network, with the $F_1$ score as the loss function. The $F_1$ score function is the harmonic average of the precision and recall of the classification results. This score function is often used whenever the number of positive samples is small. It is frequently used in the field of information retrieval, as the number of relevant documents is much smaller than the total number of documents. In our scenario, where the number of positive (buggy) class samples is much smaller than the number of negative samples, using this score function as the loss function will improve the learning precision. 

Furthermore, we applied negative sub-sampling in the training phase of the neural network. This technique is implemented in the training phase of the neural network. The neural network is trained in 20 iterations. In each iteration all the positive samples and a subset of the negative samples is used for training. The negative subset is a random subset of the total negative samples with the size proportional to the size of positive samples set.

\subsubsection{Proposed algorithm}
\label{proposed_algorithm}
\begin{figure}[t]
\centering
\includegraphics[width=3.6in]{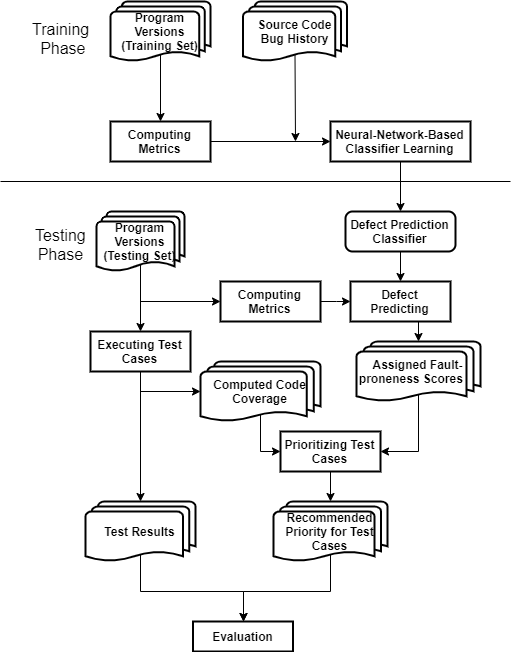}
\caption{Overview of the proposed test case prioritization algorithm}
\label{test_prioritization_algorithm}
\end{figure}

Figure~\ref{test_prioritization_algorithm} shows a big picture of our proposed TCP algorithm. As it can be seen in the figure, this algorithm starts with a training phase in which a defect prediction classifier is learned. In the training phase, the historical data regarding the previous bugs in the source code (referred to as \textit{Source code bug history} in Figure~\ref{test_prioritization_algorithm}) is utilized to train a defect prediction model. This model is then used to predict fault proneness in the current or later versions of the source code. More specifically, for each version of the source code history, classes reported to be buggy are marked as positive samples, and all other classes as negative samples.

In the next step, for each program version in the testing set, source code and historical metrics are computed and then used by the defect prediction classifier to assign a fault-proneness score to each code unit. Moreover, the code coverage of the test cases recorded in previous executions of the program are fetched. Next, the test coverage and assigned fault-proneness score is used to prioritize the test cases and achieve a recommended priority order. Note that in this phase different strategies could be used. For the evaluation of the algorithm, the actual test results and recommended priority for the test cases are used to compute the APFD score.

\subsubsection{Modified strategies}
When the test coverage is computed, various strategies can be used to prioritize the test cases. For example, the traditional strategies reviewed in Section~\ref{traditional_strategies} could be used for this purpose. As it was explained in Section~\ref{proposed_algorithm}, another option is to use a modified version of these strategies which, in addition to the test coverage data, also takes into account the code units fault-proneness score.

In this paper, we propose a method to obtain such modified strategies. In this manner, we can substitute the traditional coverage (as defined in Equation~\ref{total_coverage_equation}) with the \textit{FaultBasedCover} (introduced in Equation~\ref{weighted_cover}) and then perform the traditional strategies using this definition of test coverage. 
We call the resulting strategies the \textit{modified strategies}, which will be described in the following subsections.
\subsubsection{Modified Additional Strategy}
Algorithm~\ref{mod_add_strategy} shows the algorithm resulted from applying the proposed modifications to the additional strategy. Before executing Algorithm~\ref{mod_add_strategy}, a neural network must be learned using the set of recorded bugs and previous versions of the source code (See Section~\ref{proposed_dp_method}). The resulting model will be used as the input \textit{DPModel} in the algorithm.


\begin{algorithm}[!htb]
\caption{Proposed modified additional algorithm}\label{mod_add_strategy}
\begin{algorithmic}[1]
\Inputs{\textit{Metrics:} computed metrics \\ \textit{DPModel:} the learned defect prediction model \\ \textit{Cover:} the coverage matrix of the test suite}
\Outputs{\textit{Priority:} the recommended order}
\State $P_{dp} \gets defectPrediction(Metrics, DPModel)$
\State $n \gets numberOfRows(Cover)$
\State $m \gets numberOfColumns(Cover)$
\For {each $j (1 \leq j \leq m)$}
    \State $Prob[j] \gets  P_0 + (1-P_0) \times P_{dp}(j) $
\EndFor
\For {each $i (1 \leq i \leq n)$}
    \State $Total[i] \gets 0$
    \For {each $j (1 \leq j \leq m)$}
        \State $Total[i] \gets Total[i] + Prob[j] \times Cover[i,j]$
    \EndFor
\EndFor
\For {each $i (1 \leq i \leq n)$}
    \State $Selected[i] \gets false $
\EndFor
\For {each $i (1 \leq i \leq n)$}
    \State $k \gets 1 $
    \While {$Selected[k]$}
        \State $k \gets k+1$
    \EndWhile
    \State $sum \gets 0$
    \For {each $j (1 \leq j \leq m)$}
        \If {$Cover[k,j]$}
            \State $sum \gets sum + Prob[j]$
        \EndIf
    \EndFor
    \For {each $l (k+1 \leq l \leq n)$}
        \If {$\neg Selected[l]$}
            \State $s \gets 0$
            \For {each $j (1 \leq j \leq m)$}
                \State $s \gets s + Prob[j] \times Cover[l,j]$
            \EndFor
            \If {$(s \geq sum)\lor$ \State $(s = sum \land Total[l] \ge Total[k])$}
                \State $sum \gets s$
                \State $k \gets l$
            \EndIf
        \EndIf
    \EndFor
    \State $Priority[i] \gets k$
    \State $Selected[k] \gets true$
    \For {each $j (1 \leq j \leq m)$}
        \State $Prob[j] \gets max(Prob[j]-Cover[k,j], 0)$
    \EndFor
\EndFor
\end{algorithmic}
\end{algorithm}

Algorithm~\ref{mod_add_strategy} consists of three major steps. The first step of this algorithm (line 1) runs the defect prediction method and produces $P_{dp}$ array. The $j$th element of this array represents the fault-proneness score assigned to the $j$th code unit. In the second step (which is performed in lines 2-9), the value of FaultBasedCover is calculated for each test case. In the third step (shown in lines 10-32), test cases are ordered using a greedy algorithm. According to this greedy algorithm, in each step, the test case with the highest fault based coverage over the uncovered code area is chosen as the next test case. The rest of the algorithm is similar to the traditional additional strategy explained in Section~\ref{trad_add_strat}.

In the implementation of the additional method the following details were considered:
\begin{enumerate}
\item Whenever there is a tie between multiple test cases with the same additional coverage, the test case with bigger total coverage is chosen to break the tie. In case both the total and additional coverages are equal, the test case which appears first in the test suite is chosen.
\item The additional coverage of all test cases is computed at the beginning and whenever a test case is chosen, its coverage is decreased from all units.
\end{enumerate}

The first, second, and third steps of this algorithm run in time complexity $O(mf)$, $O(nm)$, and $O(n^2m)$, respectively, where n is the number of test cases, m shows the number of code units and f determines the number of features which have been used for defect prediction. Therefore, the whole algorithm runs in the time complexity of $O(mf+n^2m)$.

\subsubsection{Modified total strategy}
The modified total strategy is developed in a process similar to the modified additional strategy. The first and second steps of the this algorithm is exactly the same as the modified additional strategy, but the last step sorts the test cases by the FaultBasedCover values of test cases. The third step can be implemented by an efficient sort algorithm so the time complexity of third step is $O(n\log{n})$. As a result, the whole algorithm runs in the time complexity of $O(mf+nm+n\log{n})$.

\subsubsection{Time complexity comparison}
The time complexity analysis of the modified strategies, is summarized in Table~\ref{tab:time_complexity}. The time complexity added from the proposed modification in both the total and additional strategies is $O(mf)$. The number of features used in our implementations is constant ($f=104$), and therefore $O(mf)$ is equivalent to $O(m)$. Therefore, the computational complexity of the modified algorithms is equivalent to their traditional counterparts (as shown in Table~\ref{tab:time_complexity}). Experimental observations also show a very small difference (for all projects less than 40 milliseconds) of execution time between the traditional and modified algorithms.

\begin{table}[!htb]
\captionsetup{font=normalsize}
\centering
\caption{Time complexity of the traditional and modified strategies}
\label{tab:time_complexity}
\begin{tabular}{M{1.5cm}|M{3cm}|M{3cm}}
\hline
	\textbf{Strategy} & \textbf{Traditional} & \textbf{Modified}  \\ \hline \hline
	Additional & $O(n^2m)$ & $O(n^2m+mf)$
	
	$= O(n^2m)$ \\ \hline
	Total & $O(mn+n\log{n})$ & $O(mn+n\log{n}+mf)$
	
	$= O(mn+n\log{n})$ \\ \hline
\end{tabular}
\end{table}

\section{Empirical study}
\label{empirical_study_sec}
In this section, we explain our empirical study and discuss the results of our experiments.

\subsection{Research Questions}
\label{research_questions}
In our empirical study, we answer the research questions, presented in the introduction of this paper. These research questions can be more precisely stated as follows:
\begin{itemize}[label={}]
    \item {\bf RQ1:} How does the modified additional TCP strategy compare to the traditional additional TCP strategy in terms of APFD?
    \item {\bf RQ2:} How does the modified total TCP strategy compare to the traditional total TCP strategy in terms of APFD?
    \item {\bf RQ3:} How does the value of $P_0$ affect the effectiveness of the modified strategies?
\end{itemize}

\subsection{Subjects of study}
\label{subjects_of_study}
Among previous TCP researches, some studies have used datasets with real bugs and some have used mutation analysis methods to artificially create buggy versions of the code.

Artificially created bugs are created using a random process of injecting bugs in the source code; hence, these bugs do not represent the behavior of a software development team. In this study, we want to propose a method for TCP in order to find the bugs made by the development team as soon as possible. Therefore, we limited our study to a dataset with real software bugs.

Conclusively, the proposed method must be evaluated on projects with the following properties:
\begin{enumerate}
\item
The project must contain a test suite that is large enough to be used for TCP.
\item
The bugs of the project must be recorded during a long period of time of the development process of the project, leading to a bug database with enough number of bugs.
\item
The version control asset of the project must contain the faulty versions of the software that result in failing test cases. Moreover, the failing test cases must be identifiable.
\end{enumerate}

Just et al. have collected a dataset, namely Defects4J \cite{just2014defects4j}, which has the mentioned properties. Considering the mentioned criteria, Defects4J is one of the few datasets that can be used for this purpose. In its initial published version, Defects4J provided a recorded bug history of five well-known open source Java projects which contain a considerable number of test cases, summarized in Table~\ref{tab:def}. As it represents a completely real project development history, hopefully the results will be practically significant.

Per each recorded bug, Defects4J provides two versions of the project. First, a faulty version which contains the bug and one or more failing test cases identifying the bug. Second, another version with the bug fixed and no failing test cases. Defects4J localizes and isolates each bug fix, such that the difference between the buggy version and the fixed version is a single git commit containing only bug fix changes. This helps us to locate buggy classes in each version of the source code.

\begin{table}[!htb]
\captionsetup{font=normalsize}
\centering
\caption{Projects included in Defects4J initial version}
\label{tab:def}
\begin{tabular}{c|c|c|c}
\hline
	\textbf{Identifier} & \textbf{Project name} & \textbf{Bugs} & \textbf{Test classes} \\ \hline \hline

	Chart & JFreechart & 26 & 355\\ \hline
	Closure & Closure compiler & 133 & 221 \\ \hline
	Lang & Apache commons-lang & 65 & 112\\ \hline
	Math & Apache commons-math & 106 & 384\\ \hline
	Time & Joda-Time & 27 & 122\\ \hline \hline
	\textbf{Sum} & \textbf{-} & \textbf{357} & \textbf{1194}\\ \hline
\end{tabular}
\end{table}

\subsection{Defects4J+M: The created dataset}
Although Defects4J provides an appropriate dataset of real bugs, we still need to perform some calculations on it in order to use the data to evaluate our proposed approach. The result of these calculations is an extension of the Defects4J dataset which contains computed test coverages and source code metrics for each version included in Defects4J. The new dataset can be used by researchers in various software engineering fields, such as software repair, bug prediction, software testing, and fault localization. We made this dataset, called Defects4J+M, publicly available on GitHub\footnote{\url{https://github.com/khesoem/Defects4J-Plus-M}}.

The tests in the Defects4J projects are mainly written using the JUnit framework \cite{gamma1999junit}. In order to find test coverages we executed all test methods and measured the dynamic statement coverage. Method level was chosen as the partitioning level of code into units and JaCoCo \cite{hoffmann2011eclemma} library was used for code coverage analysis. This library is intended for easy integration with various development tools.

In this paper, we use a combination of static and process metrics for feature extraction. To extract static metrics we used the free version of SourceMeter (a tool for static code analysis). SourceMeter supports five major groups of metrics\footnote{The detailed description for SourceMeter metrics is published in its user guide page.} \cite{sourcemeter}. All the metrics were computed at class level. In addition to metrics computed using SourceMeter, we also computed the number of developers and changes per file and the number of previous bugs for each class. Table~\ref{tab:features} contains the details of each feature group. The row referred as \textit{Bug label} is the machine learning label used to train the neural network. 

In order to use the computed metrics for defect prediction, we stored them in a vector that is used as the input feature vector by the defect prediction algorithm.
\begin{table*}[!htb]
\captionsetup{font=normalsize}
\centering
\caption{Defect prediction features} \label{tab:features}
\begin{tabular}{|M{0.3cm}|M{1cm}|M{2cm}|M{5cm}|M{1cm}|M{7cm}|}
\hline
\textbf{\#} & \textbf{Feature type} & \textbf{Category} & \textbf{Definition} & \textbf{Feature count} & \textbf{General Items} \\ \hline \hline
1 & Input & Source code metrics & Used to quantify different source code characteristics & 52  & Cohesion metrics,  Complexity metrics,  Coupling metrics,  Documentation metrics,  Inheritance metrics,  Size metrics \\ \hline 
2 & Input & Clone metrics & Used to Identify the number of type-2 clones (same syntax with different variable names) & 8 & Clone Classes,  Clone Complexity,  Clone Coverage,  Clone Instances,  Clone Line Coverage,  Clone Logical Line Coverage,  Lines of Duplicated Code,  Logical Lines of Duplicated Code  \\ \hline 
3 & Input & Coding rule violations & Used for counting coding violation rules & 42  & Basic Rules,  Brace Rules,  Clone Implementation Rules,  Controversial Rules,  Design Rules,  Finalizer Rules,  Import Statement Rules,  J2EE Rules,  JUnit Rules,  Jakarta Commons Logging Rules,  Java Logging Rules,  JavaBean Rules,  Naming Rules,  Optimization Rules,  Security Code Guideline Rules,  Strict Exception Rules,  String and StringBuffer Rules,  Type Resolution Rules,  Unnecessary and Unused Code Rules  \\ \hline 
4 & Input & Git metrics & Used to count the number of committers and commits per file (these metrics could not be computed for inner classes) & 2  & Committers count, Commit counts \\ \hline 
5 & Output & Bug label & Label that shows this file is buggy in this version of the project or not & 1 & IsBuggy \\ \hline
\end{tabular} 
\end{table*}

The source code metrics and test coverages were computed using a machine with two 10-core Intel CPUs (Intel(R) Xeon(R) CPU E5-2690 v2 @ 3.00GHz) with 256GB RAM. It took more than 100 hours for this computer to finish the calculations. In addition to the large amount of resources and time that had to be dedicated to these calculations, we also faced some difficulties while creating the dataset that make it reasonable for us to release the dataset publicly so that other researchers can use it without facing the same difficulties. A short list of these difficulties is as follows:
\begin{itemize}
\item   In some projects, different program versions had to be built using different building tools, such as Ant, Maven, or Gradle.
\item   In some projects, different program versions used different Java Development Kit (JDK) versions.
\item   Computing metrics and test coverages took very long for some versions. Therefore, we had to do the calculations in parallel in order to finish them in a reasonable length of time.
\item   A few versions of some projects could not be compiled. Moreover, the process of calculating metrics or test coverages for some projects took too long. We had to recognize such cases and ignore them if they could not be fixed.

\end{itemize}

\subsection{Experimental procedure}
\label{experimental_procedure}
The experiment consists of running the proposed method, the total and the additional prioritization methods on the projects of the Defects4J+M dataset. In order to create the defect prediction model for the $i$th version of a project, the procedure explained in Section~\ref{proposed_dp_method} is performed using the data from the $1$st to $(i-1)$th versions of the same project. The resulting model was then used for TCP. Since creating the defect prediction model requires the data from, at least, a minimum number of buggy versions, we created the model only for the more recent versions of each project. In this regard, the evaluation is done over the last 13, 33, 50, 14, and 50 versions of the Chart, Lang, Math, Time, and Closure projects, respectively.

The source code and usage instructions of the methods implemented in this paper are made publicly available on a GitHub project\footnote{\url{https://github.com/mostafamahdieh/FaultPronenessBasedTCP}}. This package contains detailed instructions on usage and replicating the results of this paper in multiple steps.

The defect prediction neural network is implemented using Python language and \textit{Keras} and \textit{scikit-learn} machine learning libraries. The TCP algorithm is also implemented with Python language using \textit{NumPy} and \textit{pandas} libraries.

Table~\ref{tab:dp_performance} shows some properties of the learning process. The column \textit{Evaluation versions} shows the number of versions that the project is evaluated on. As the number of input samples varies in different versions, the minimum and maximum number of input samples used in the learning process for each project is shown in column \textit{Min input samples} and \textit{Min input samples}

The neural network performance can be measured by counting the number of predicted bugs using the neural network. More exactly, for each version of the project, if a bug is predicted using the defect prediction model, this value is incremented. A bug is considered to be predicted, if a buggy class corresponding to its bug-fix is assigned fault-proneness value of more than 0.1. This value is displayed in column \textit{Predicted bugs} of Table~\ref{tab:dp_performance}.


\begin{table}[!htb]
\captionsetup{font=normalsize}
\centering
\caption{Performance of defect prediction}
\label{tab:dp_performance}
\begin{tabular}{M{1cm}|M{1.5cm}|M{1.4cm}|M{1.4cm}|M{1.2cm}}
\hline
	\textbf{Project} & \textbf{Evaluation versions} & \textbf{Min input samples} & \textbf{Max input samples} & \textbf{Predicted Bugs} \\ \hline \hline
	Chart & 13 & 11121 & 21763 & 3/13 \\ \hline
	Closure & 50 & 57570 & 102459 & 25/50 \\ \hline
	Lang & 33 & 7128 & 13863 & 16/33 \\ \hline
	Math & 50 & 33443 & 76804 & 22/50 \\ \hline
	Time & 14 & 6330 & 10154 & 5/14 \\ \hline \hline
	\textbf{Overall} & \textbf{160} & \textbf{6330} & \textbf{102459} & \textbf{71/160} \\ \hline
\end{tabular}
\end{table}


\section{Results}
\begin{figure*}[!htb]
\centering
\centerline
{\subfloat[Math]{\includegraphics[width=1.4in]{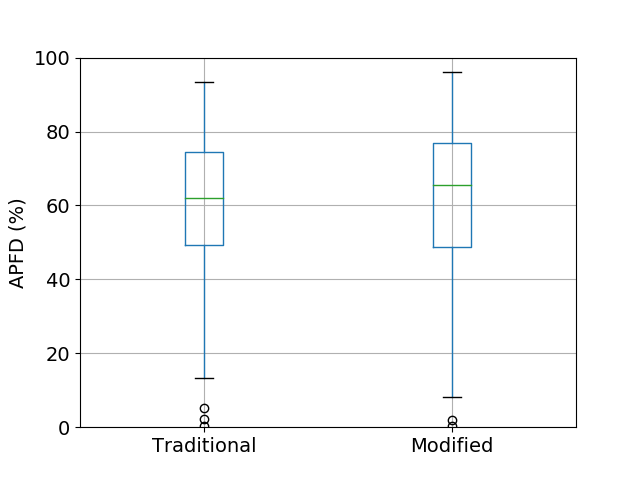}}
\subfloat[Chart]{\includegraphics[width=1.4in]{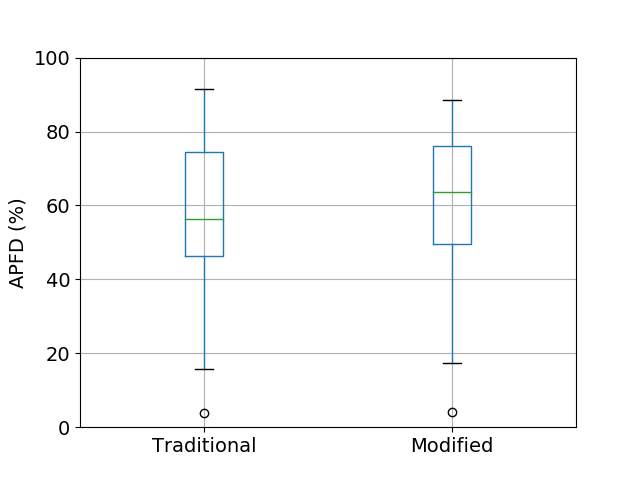}}
\subfloat[Time]{\includegraphics[width=1.4in]{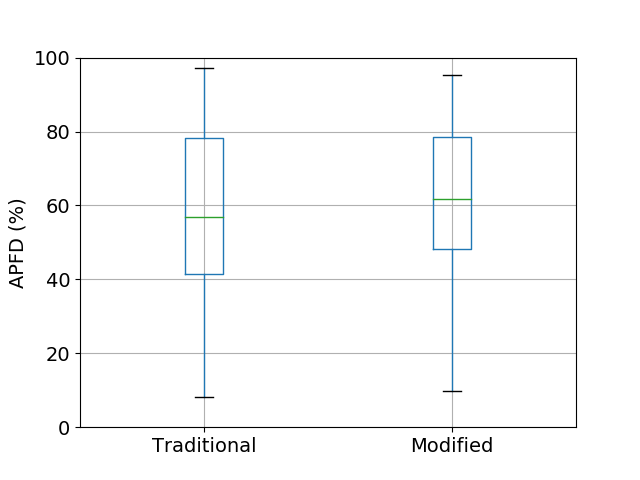}}
\subfloat[Lang]{\includegraphics[width=1.4in]{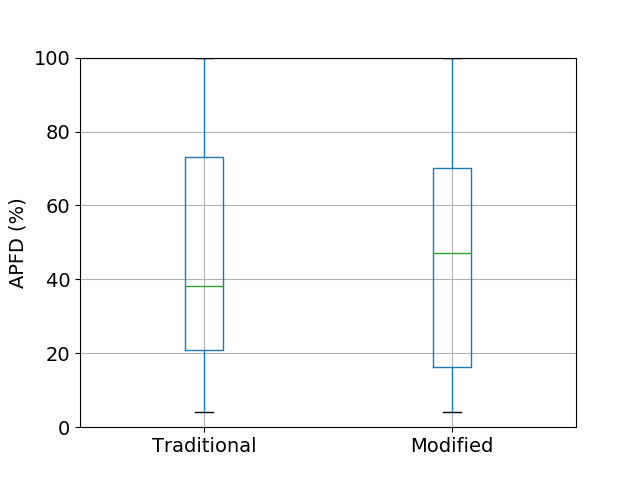}}
\subfloat[Closure]{\includegraphics[width=1.4in]{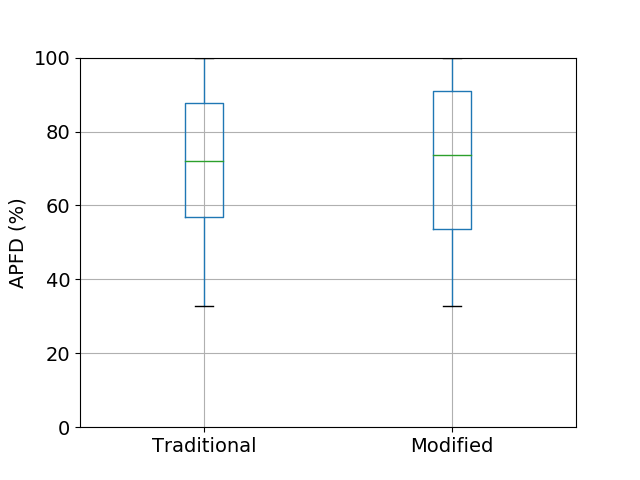}}}
\caption{Evaluation results of traditional and modified additional TCP strategies (RQ1)} \label{mod_vs_add}
\end{figure*}

\label{section_results}
In this section, we present the results of our empirical study. In this regard, we provide the experimental results to answer the research questions raised in Section~\ref{research_questions}.

\subsection{RQ1 and RQ2: Comparing modified strategies with traditional strategies}
In order to answer RQ1 and RQ2, we computed the APFD score of traditional total and additional strategies as well as that of their modified versions. For each project of study, the APFD value is extracted over all the versions considered for evaluation in Section~\ref{experimental_procedure}. Our preliminary experiments showed that our approach works the best when $P_0$ is set to $0.3$. Therefore, we compare the modified strategies with traditional ones using this setting. The effect of changing the value of $P_0$ is further discussed in Section~\ref{effect_of_p0}.

\begin{figure*}[!htb]
\centering
\centerline
{\subfloat[Math]{\includegraphics[width=1.4in]{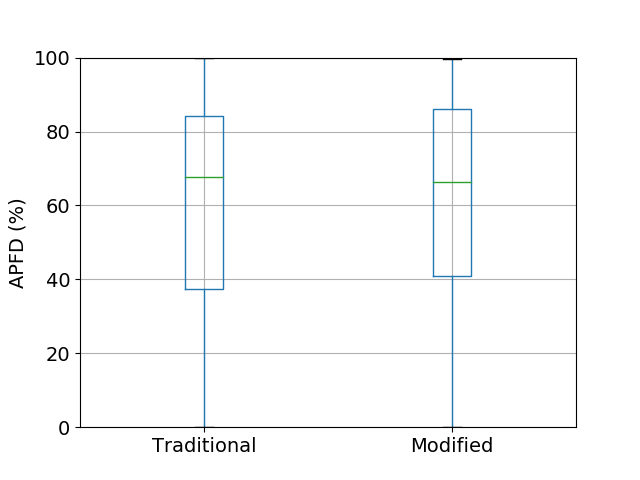}}
\subfloat[Chart]{\includegraphics[width=1.4in]{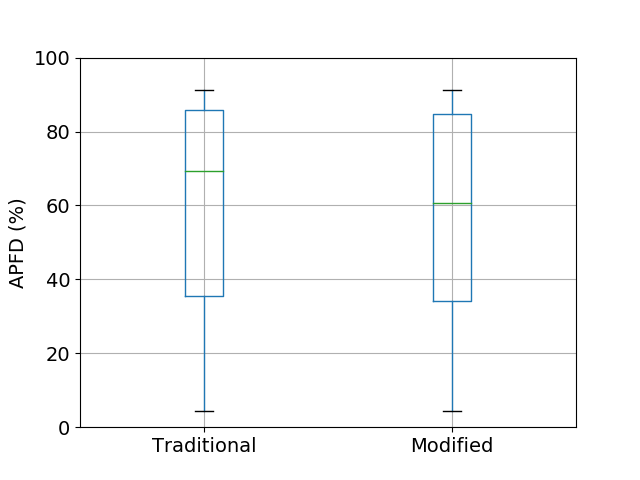}}
\subfloat[Time]{\includegraphics[width=1.4in]{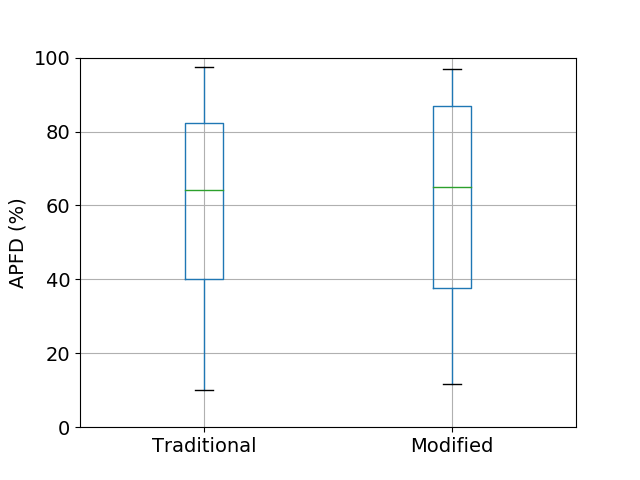}}
\subfloat[Lang]{\includegraphics[width=1.4in]{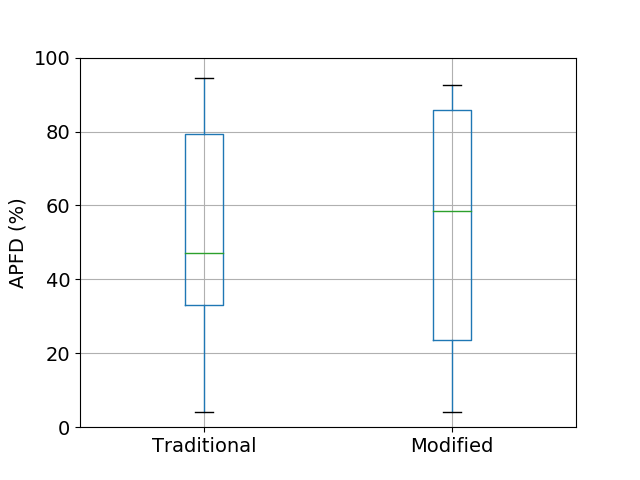}}
\subfloat[Closure]{\includegraphics[width=1.4in]{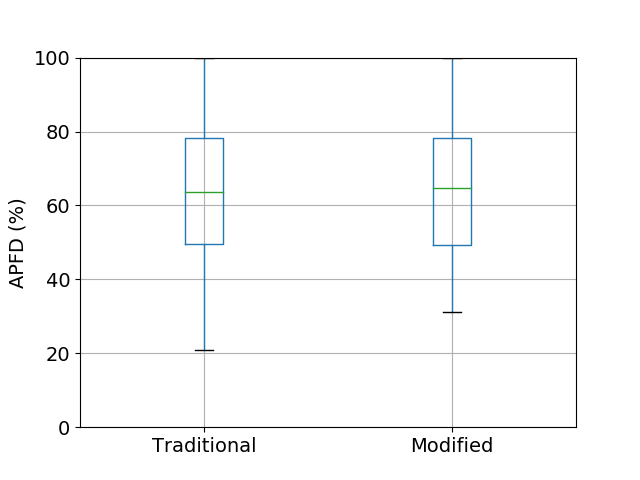}}}
\caption{Evaluation results of traditional and modified total TCP strategies (RQ2)} \label{mod_vs_total}
\end{figure*}

The box plots in Figure~\ref{mod_vs_add} and Figure~\ref{mod_vs_total} represent the APFD score of the modified and traditional versions of additional and total strategies on each subject of study, respectively. Table~\ref{apfd_scores} also shows the mean APFD score of each strategy on each project. The \textit{improvement} column is the mean improvement of the APFD value of the modified strategy compared to the traditional strategy, defined using Equation~\ref{improvement_formula}:

\newcommand\APFD{A\!P\!F\!D}

\begin{equation}
\label{improvement_formula}
\mathit{impr} = \frac{1}{|\textit{Versions}|} \sum\limits_{v\in \textit{Versions}} {\frac{\APFD^{\mathit{Mod}}_v-\APFD^{\mathit{Trad}}_v}{\APFD^{\mathit{Trad}}_v}}
\end{equation}

Where $\APFD^\mathit{Mod}_v$ and $\APFD^\mathit{Trad}_v$ denote the APFD of the modified and traditional algorithms, $\mathit{Versions}$ represents the set of all versions under evaluation and $\mathit{impr}$ stands for the improvement value.

This value can be used as a measure to show the magnitude of improvement of the modified algorithms with respect to the traditional algorithms. Note that this value is not equal to the difference of the value of the modified column (mean APFD value of the modified strategy) to the traditional column (mean APFD value of the total strategy).

As it can be observed in Table~\ref{apfd_scores}, the mean APFD scores of modified strategies are superior to that of traditional strategies in most occasions and especially, in the overall case. This can be seen numerically in the improvement column, which is positive in most of the cells. The overall improvement for the five projects is 4.63\% and 3.60\% for of the additional and total strategy respectively.

Moreover, we performed a Wilcoxon signed-rank test \cite{wilcoxon1992individual} ($p$-value $< 0.05$) to make sure that our results are statistically significant. The null hypothesis is that there is no significant difference in the performance of the modified strategies with respect to their traditional counterparts. The results of this test demonstrate that:

\begin{enumerate}
\item There is overally a statistically significant difference between the modified and traditional additional strategies ($p$-value $ = 0.00268$). This means that the modified additional strategy significantly performs better than the traditional one (RQ1).
\item There is no significant difference between the modified and traditional total strategies (RQ2, $p$-value $= 0.406$). Furthermore, Figure~\ref{mod_vs_total} confirms this result since it shows that the modified total strategy does not outperform the traditional total strategy in terms of the median APFD value (note that the mean value of APFD is not shown in the box plot). In three of the projects (Math,  Time, and Closure), the differences between the median APFD of the modified and traditional strategies are negligible, on the Chart project the traditional strategy is superior and only on the Lang project the modified strategy is superior.
\end{enumerate}

\begin{table*}[!htb]
\captionsetup{font=normalsize}
\centering
\caption{Average APFD scores of each TCP strategy}
\label{apfd_scores}
\begin{tabular}{c|c|c|c|c|c|c}
 \hline
	\multirow{2}{*}{\textbf{Subject}} &
	\multicolumn{3}{c|}{\textbf{Additional Strategy}} & \multicolumn{3}{c}{\textbf{Total Strategy}} \\ 
	\cline{2-7}
                        & \texttt{Traditional} & \texttt{Modified} & \texttt{Improvement}  & \texttt{Traditional} & \texttt{Modified} & \texttt{Improvement} \\ \hline
\hline
Chart                       & 55.47\%                & \textbf{57.84\%} &       5.95\%      & \textbf{58.59\%}                  & 57.25\% & -2.68 \%                \\ \hline
Closure                       & \textbf{71.06\%}                & 71.01\% &  -0.14   \%    & 64.13\%                  & \textbf{64.59\%} & 1.44\%                   \\ \hline
Lang                       & 45.76\%                & \textbf{46.92\%} &  7.41\% & 53.80\% & \textbf{54.51\%} & 6.95\% \\ \hline
Math                       & 59.43\%                & \textbf{60.78\%} & 4.37\% & 60.84\%                  & \textbf{61.88\%} & 9.32\%                   \\ \hline
Time                       & 57.48\%                & \textbf{58.65\%}      & 5.57\%          & 59.28\%                  & \textbf{60.15\%}         & 2.99\% \\ \hline \hline
Overall                     & 59.54\%                & \textbf{60.50\%}             & 4.63\% & 60.02\%                  & \textbf{60.60\%} & 3.60\%                \\ \hline
	
\end{tabular}
\end{table*}


\subsection{RQ3: Investigating the effect of changing \texorpdfstring{$P_0$}{P0} on the effectiveness of modified strategies}
\label{effect_of_p0}

As it was mentioned in Section~\ref{modified_coverage}, Equation \ref{proposed_probability_function} estimates the fault-proneness probability as a linear combination of a constant value (1) and the fault-proneness score assigned by the defect prediction model ($P_{dp}$), using the parameter $P_0$ or equivalently $1-P_0$ denoted as $C_{dp}$. 

\begin{figure*}[!htb]
\centering
{\subfloat[Math]{\includegraphics[width=1.4in]{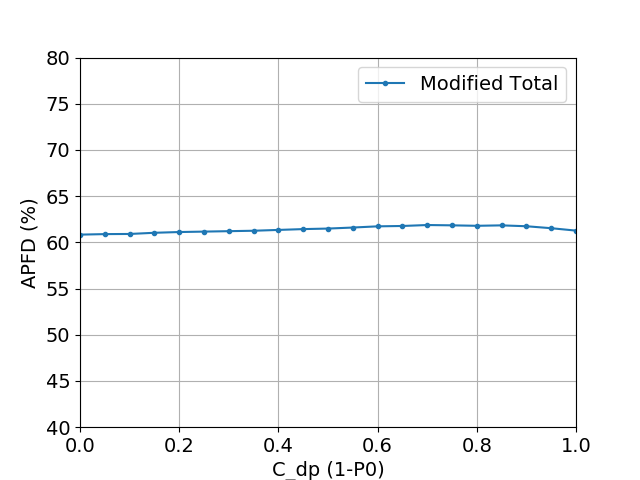}}
\subfloat[Chart]{\includegraphics[width=1.4in]{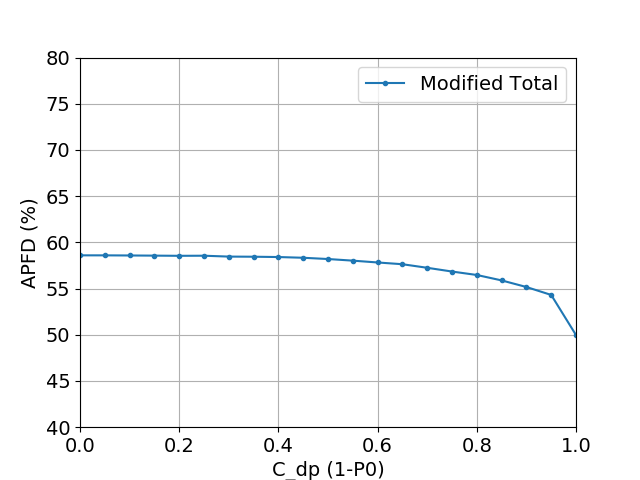} \label{mod_vs_p0_total_chart}}
\subfloat[Time]{\includegraphics[width=1.4in]{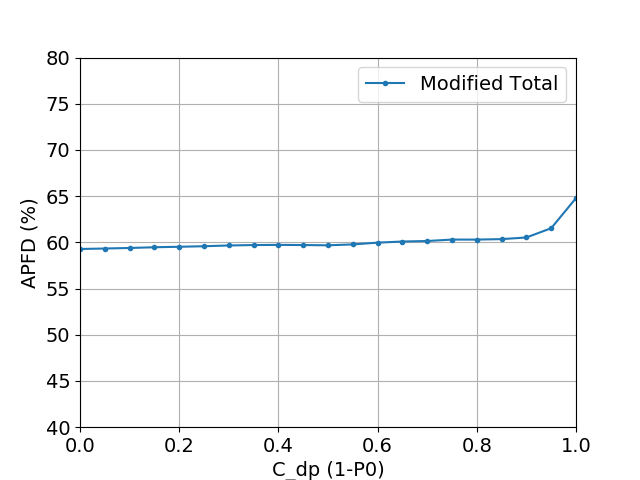}}
\subfloat[Lang]{\includegraphics[width=1.4in]{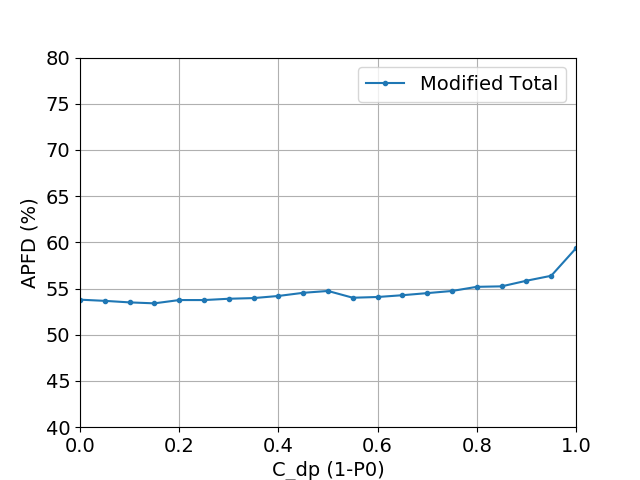}}
\subfloat[Closure]{\includegraphics[width=1.4in]{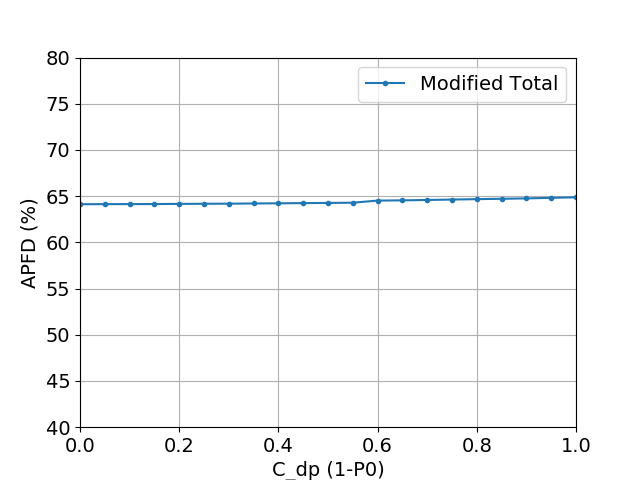}}}
\hfill
\centerline
{\subfloat[Math]{\includegraphics[width=1.4in]{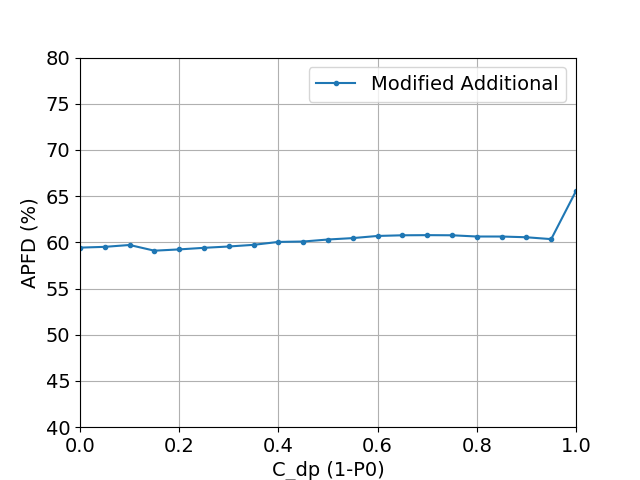}}
\subfloat[Chart]{\includegraphics[width=1.4in]{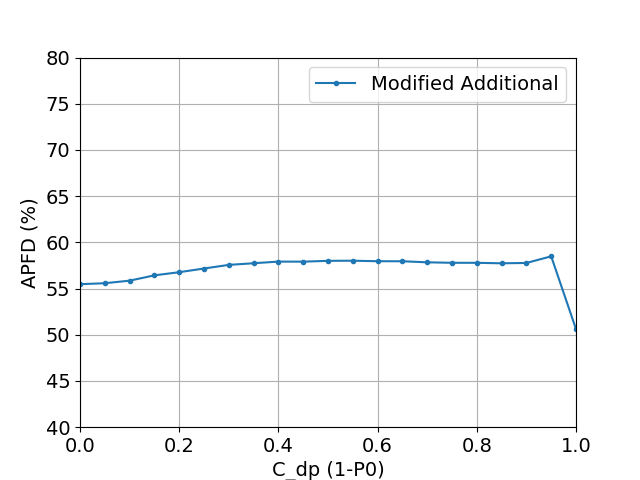}}
\subfloat[Time]{\includegraphics[width=1.4in]{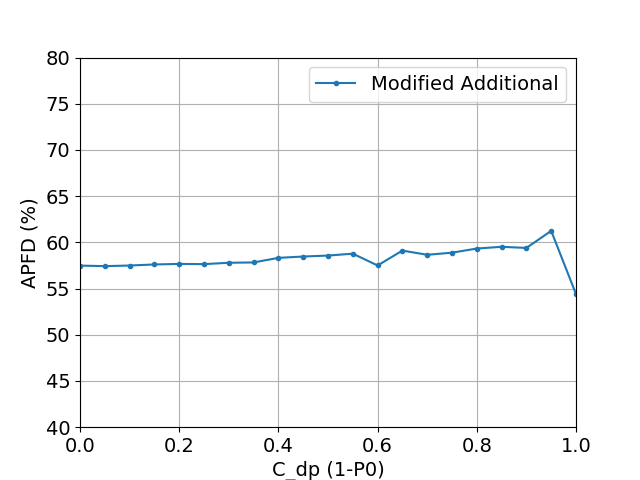}}
\subfloat[Lang]{\includegraphics[width=1.4in]{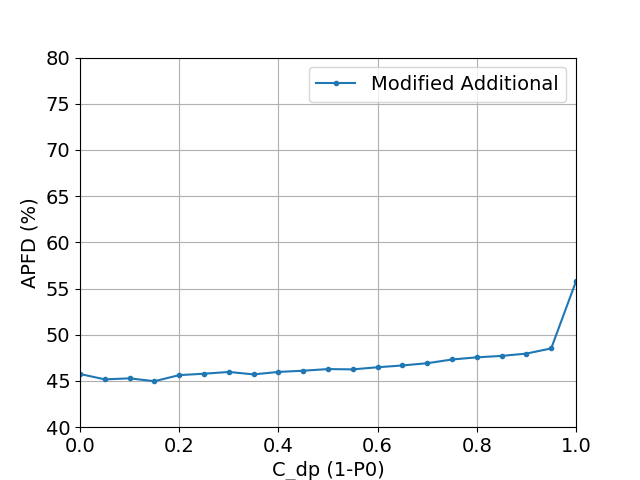}}
\subfloat[Closure]{\includegraphics[width=1.4in]{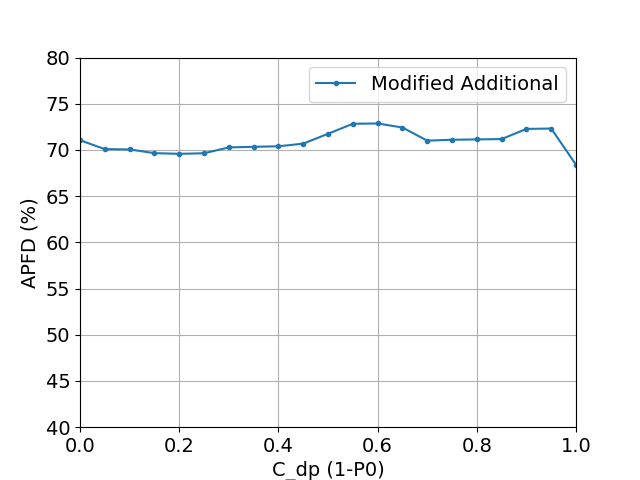}}}
\caption{The APFD performance of the modified strategies with respect to varying $C_{dp}$ (RQ3)} \label{mod_vs_p0}
\end{figure*}

The relation between the mean APFD values versus $C_{dp}=1-P_0$ is plotted in Figure \ref{mod_vs_p0}. Each curve in this figure is showing the performance (APFD) of a modified method on a specific project. To observe the behavior of the modified strategies, the value of $P_0$ is varied in the range $[0,1]$. This shows how the value of APFD changes in response to changing the value of $P_0$.

The value of $C_{dp}$ can be set by the practitioner to an appropriate value regarding the project conditions. From one point of view, $C_{dp}$ tunes the amount of confidence to the predicted fault-proneness values. In the extreme cases of the interval $[0, 1]$, setting $C_{dp}$ to one gives full confidence to the defect prediction method and setting $C_{dp}$ to zero ignores the defect prediction and only takes the coverage into account. When the project provides more prior knowledge (e.g. bug history size), $C_{dp}$ can be set to a higher value to increase the impact of this knowledge; otherwise, it should be set to lower values.

Figure~\ref{mod_vs_p0} shows the relation between the APFD and $C_{dp}$ in modified total and modified additional strategies. Note that when the value of $C_{dp}$ is set to $0$, the modified strategy works the same as the traditional strategy. Therefore, in each curve, on the points that are higher than the most left point of the curve, the modified strategy is working better than the traditional one.

As it can be observed in Figure~\ref{mod_vs_p0}, the curves show an increasing trend in most cases; therefore, it can be claimed that the modified strategy performs better when the fault-proneness score is more taken into account. An exception to this claim is Figure~\ref{mod_vs_p0_total_chart}. This Figure shows the effect of changing $C_{dp}$ on the APFD value in the modified total strategy for the Chart project. We believe that this exception occurs because of the low number of bugs (26) recorded for the Chart project which, in turn, causes an inaccurate defect prediction model.

\section{Discussion}
\label{section_discussion}
\subsection{Practical Considerations}
The proposed modification method relies on the success of the defect prediction phase. Therefore, practically this method will fail in case the requirements for successful defect prediction, such as a large enough history of bugs, are not met. Moreover, the bug history must contain real bugs related to a single software development team, not artificial bugs or bugs created by mutants.

The time required for executing the defect prediction phase is almost the same as the time needed for running the additional TCP strategy (in terms of order of magnitude). However, we can avoid executing the defect prediction phase for each prioritization task because the same defect prediction model created for a recent project version can also be used for the last version.

\subsection{Threats to validity}
\textit{Internal Validity.} As mentioned in Section~\ref{proposed_dp_method} our proposed method estimates the fault-proneness of the methods using defect prediction on classes and extrapolates the fault-proneness values assigned to classes to estimate the fault-proneness of methods. Using this approach results in more false positive instances for the prediction of bugs (i.e. falsely reporting a method as being buggy). In this regard, we measured the number of classes being reported as buggy, and our observations show that only a small ratio of classes are reported as buggy which means the ratio of false positives would be low as well. Hopefully, a small amount of false positives would not be harmful for our results.

\textit{External Validity.} Our subject projects are all implemented in the Java language; therefore, the results might differ in projects developed using other languages. Furthermore, all the projects used in our empirical study are popular open-source projects and contain large test suites while projects that use our approach in the future might not have these features. We conducted the statistical test in order to make sure that our results can be generalized with confidence; however, we still have to evaluate our approach using projects with different languages and characteristics in the future to ensure the results are generalizable.

\section{Related work}
\label{related_work_sec}
There have been many methods proposed for the problem of test case prioritization. Among methods proposed for TCP, the largest category is the category of the coverage-based methods. These methods rely on the assumption that choosing test cases with larger coverage leads to more effective fault detection. Therefore, these methods attempt to prioritize the test cases in an order that has the most coverage in the least number of executed test cases. Coverage-based TCP methods inherently solve an optimization problem, which is related to the set cover problem \cite{offutt1995‎} and is proved to be NP-hard \cite{Li2007}. Therefore, there is no polynomial time algorithm for computing an optimal solution to the coverage-based TCP problem, and algorithms proposed for the coverage-based TCP problem are heuristic methods to solve this problem.

Coverage measurement is done in two main categories: Dynamic coverage and static coverage. Dynamic coverage is  measured by executing the test cases and auditing the execution trace of each test case. Static coverage is an approximate estimation of dynamic coverage, measured by analyzing the static structure of the source code. Dynamic code coverage is widely used in many existing TCP studies; however, static code coverage, which is the coverage estimated from static analysis, has also been studied for TCP \cite{mei2012static, zhou2017impact, zhang2009prioritizing}.

Each coverage-based TCP method consists of two main building blocks: first, the coverage criteria used to measure coverage and second, the strategy used to take into account the measured test case coverage for TCP. Regarding the first building block, various coverage criteria have been applied for coverage-based TCP. The early studies in TCP used the statement coverage, branch coverage \cite{rothermel1999test}, and method coverage \cite{elbaum2000prioritizing}. After that multiple other criteria have been proposed \cite{jones2003test, kovacs2009optimal, fang2014similarity}. Fang et al. compare major existing logic and fault-based coverage criteria \cite{fang2012comparing}. Their main conclusion is that criteria with fine-grained coverage information, MC/DC and fault based logic coverage criteria, have better fault detection capability. Elbaum et al. incorporate \textit{fault index}, a metric calculated using a combination of multiple measurable attributes of the source code \cite{elbaum1999evolution}. There proposed method tracks variations of fault index over regressions and prioritizes test cases which run code with higher fault indexes.

As for the second building block of TCP methods, the strategy/algorithm can be thought as of a method to solve the optimization problem underlying coverage-based TCP. The aim of this optimization problem is to maximize the coverage with the hope that a high test coverage results in a high fault detection rate. For instance, the total and additional techniques are simple greedy algorithms providing approximate solutions for this optimization problem \cite{rothermel1999test}. Being simple, efficient, and effective has popularized the usage of these basic methods \cite{hao2014unified}. The random strategy, which simply randomly orders the test cases is used for comparison with techniques in this research area. Li et al. \cite{Li2007, li2010simulation} have applied some well known meta-heuristic optimization algorithms including hill climbing, $K$-optimal greedy algorithms (refer to \cite{Li2007} for its definition) and genetic algorithms to solve the coverage-based TCP problem. When compared to the total and additional prioritization algorithms using different metrics, the results indicate that the additional prioritization algorithm and the 2-optimal greedy algorithm, despite their simplicity, are the most efficient techniques in the majority of cases and the differences between the additional and 2-optimal algorithm are insignificant.

Zhang et al. \cite{zhang2013bridging, hao2014unified} introduce strategies to mix  the additional strategy and the total strategy resulting in a spectrum of algorithms in between them. Their results show that the mixed strategy may outperform both the additional strategy and the total strategy in terms of APFD. Jiang et al. \cite{jiang2009} propose the random adaptive strategy, a variation of the greedy additional method with a technique used for choosing between ties. Their tie strategy technique chooses the test case farthest away from the currently selected test cases. They report that the random adaptive strategy does not improve effectiveness in terms of APFD but has better computational performance. Hao et al. \cite{hao2013adaptive} propose utilizing the intermediate output of the execution process to improve test-case prioritization. 

Hao et al. \cite{hao2016optimal} focus on the coverage-based TCP problem to evaluate how much improvement is available in this problem. In this manner, they formulate this problem as an integer linear programming (ILP) problem, to produce the optimal solution and study its empirical properties. This empirical study demonstrates that the optimal technique can only slightly outperform the additional coverage-based technique with no statistically significant difference in terms of coverage. However, the additional technique significantly outperforms the optimal solution in terms of either fault detection rate or execution time. Note that their algorithm is not computationally practical and is only designed to compare the optimal solution with the output of other algorithms. 

Some authors have leveraged similarity and information retrieval metrics between test cases and source code to prioritize test cases. Saha et al. \cite{saha2015information} introduced a new approach to address the problem of TCP by reducing it to a standard information retrieval problem such that the differences between two program versions form the query and the test cases constitute the document collection. Noor et al.  \cite{noor2015similarity} proposed a TCP approach that uses historical failure data of test cases. Their method uses similarity between test cases considering a test case as effective if it is similar to any failed test cases in the previous versions of the source code. 

Some researchers have proposed using development process information to help rank test cases. Arafeen et al. \cite{arafeen2013test} used software requirements to cluster test cases and rank them. Ledru et al. \cite{ledru2012prioritizing} proposed a method that doesn't assume the existence of code or specification and is based only on the text description of test cases, which may be useful in cases where the code coverage information is not available. Korel et al. \cite{korel2005test, korel2008application} proposed a model-based method for regression TCP, which assumes the system has been modeled using a state-based modeling language. In this method, when the modifications to the source code are made, developers identify model elements (i.e., model transitions) that are related to these modifications. Then the test suite is prioritized according to the relevance of test cases to these modifications. Some researches are focused on practical constraints in TCP, such as  time constraints \cite{zhang2009time, suri2011analyzing, do2010effects, marijan2013test, zhang2007test} and fault severity \cite{walcott2006timeaware, huang2012history}.

Engstrom et al. \cite{engstrom2010empirical} proposed utilizing previously fixed faults to choose a small set of test cases for regression test selection. Laali et al. \cite{laali2016test} propose an online TCP method that utilizes the locations of faults revealed by executed test cases in order to prioritize the non-executed test cases. Some studies such as \cite{anderson2014improving} and  \cite{engstrom2011improving}, use the idea of utilizing the history of failed regression tests to improve future regression testing phases. Kim et al. employed methods from fault localization to improve test case prioritization. Using the observation that defects are fixed after being detected, they propose that test cases covering previous faults will have lower fault detection possibility \cite{kim2010effective}. \cite{wang2017qtep} proposed quality-aware test case prioritization method (QTEP) which focuses on potentially unrevealed faults. QTEP is based on CLAMI \cite{nam2015clami}, an unsupervised defect prediction method. Their evaluation shows improvement of $7.6\%$ on average with respect to the basic prioritization methods on 7 open source Java projects. Recently, Paterson et al. \cite{paterson2019empirical} proposed a ranked-based technique to prioritize test cases based on an estimated likelihood of java classes having bugs. This likelihood is estimated using a mixture of software development history features, such as number of revisions, number of authors, and number of fixes. Their experiments show that using their TCP method reduces the number of test cases required to find a fault on average by $9.48\%$ compared with existing coverage-based strategies.

As mentioned at the beginning of this section, coverage-based methods consist of the main building blocks of the used coverage criteria and strategy. The method proposed in this paper can be categorized as an approach to modify the existing coverage criteria proposed for TCP by utilizing the fault-proneness score assigned by defect prediction methods. Among the mentioned studies, \cite{wang2017qtep} and \cite{paterson2019empirical} are the two approaches that are the closest ones to our proposed approach. However, these two studies utilize sources of information different from the sources used in the current paper. Therefore, they should not be directly compared with our approach. More specifically, QTEP (\cite{wang2017qtep}) is an unsupervised method while our method is supervised and uses the bug history data. Also the method presented by Paterson et al. \cite{paterson2019empirical} incorporates different sources of information such as test execution history, which is not used in this paper.

\section{Conclusions and future work}
\label{section_conclusion}
In this research, we introduced a novel approach to incorporate the code units fault-proneness estimations into coverage-based TCP methods. For this purpose, we proposed using the fault based coverage (introduced in Equation~\ref{weighted_cover}) instead of the traditional coverage (introduced in Equation~\ref{total_coverage_equation}). In order to investigate our proposed approach, we conducted an empirical study on 357 versions of five real-world projects included in the Defects4J dataset. Our evaluations show that traditional total and additional TCP strategies are improved when they are modified due to our proposal and the improvement of the modified additional strategy is statistically significant.

In the future, we can also take into account the test case execution results in the history of the development process in order to improve the proposed TCP methods. Moreover, cross-project defect prediction techniques \cite{zimmermann2009cross} could be used to achieve a better estimation of the code units fault-proneness. The effect of modifying other existing coverage-based TCP strategies should also be investigated. Finally, the proposed approach can also be evaluated using larger datasets with various programming languages.

\section{Acknowledgments}

The authors would like to thank Mohsen Mahdieh for his helpful comments and guidance.

\section{Conflict of interest}
To the best of the authors knowledge, there isn't  any conflict of interests to report.

\section*{References}

\bibliography{wtp-elsevier}

\end{document}